\documentclass{article}

\usepackage{microtype}
\usepackage{graphicx}
\usepackage{subfigure}
\usepackage{booktabs} % for professional tables
\usepackage{float}
\usepackage{nth}
\usepackage{multirow}

\usepackage{hyperref}

\usepackage[accepted]{icml2021}

\icmltitlerunning{Tune It Up: Music Genre Transfer and Prediction}

\begin{document}

\twocolumn[
\icmltitle{Tune It Up: Music Genre Transfer and Prediction}




\begin{icmlauthorlist}
\icmlauthor{Fidan Samet}{hu}
\icmlauthor{Oguz Bakir}{hu}
\icmlauthor{Adnan Fidan}{hu}
\end{icmlauthorlist}

\icmlaffiliation{hu}{Department of Computer Engineering, University of Hacettepe, Ankara, Turkey}

\icmlcorrespondingauthor{Fidan Samet}{fidanlsamet@gmail.com}
\icmlcorrespondingauthor{Oguz Bakir}{oguzbakir0@gmail.com}
\icmlcorrespondingauthor{Adnan Fidan}{addfinnn@gmail.com}

\icmlkeywords{Machine Learning, ICML}

\vskip 0.3in
]



\printAffiliationsAndNotice{}  % leave blank if no need to mention equal contribution

\begin{abstract}

Deep generative models have been used in style transfer tasks for images. In this study, we adapt and improve CycleGAN model to perform music style transfer on Jazz and Classic genres. By doing so, we aim to easily generate new songs, cover music to different music genres and reduce the arrangements needed in those processes. We train and use music genre classifier to assess the performance of the transfer models. To that end, we obtain 87.7\% accuracy with Multi-layer Perceptron algorithm. To improve our style transfer baseline, we add auxiliary discriminators and triplet loss to our model. According to our experiments, we obtain the best accuracies as 69.4\% in Jazz to Classic task and 39.3\% in Classic to Jazz task with our developed genre classifier. We also run a subjective experiment and results of it show that the overall performance of our transfer model is good and it manages to conserve melody of inputs on the transferred outputs. Our code is available at \url{https://github.com/ fidansamet/tune-it-up}

\end{abstract}

%-------------------------------------------------------------------------

\section{Introduction}

"Music is the food of soul." said Arthur Schopenhauer. Music is an art that appeals to everyone. People have particular interest in music of certain genres such as Jazz, Pop, and Classic. By performing music genre transfer, music from different genres can be transferred to a certain genre. Thus, new hit songs can be generated automatically, musicians can easily cover music of different genres, and the music arrangements required in this process can be reduced. By music genre prediction, the performance of music genre transfer methods can be assessed. Furthermore, recommendations can be made to the listeners based on their favorite music genres. Thus, companies can obtain more profit with useful and accurate recommendations.

Style transfer between different domains has become a hot topic in the machine learning research field. Many deep learning methods have been developed to accomplish this task. By extracting fundamental knowledge about domains with obtaining deep comprehensions, deep learning models can perform style transfer. Therefore deep generative models like Generative Adversarial Networks (GAN) \cite{goodfellow2014generative} perform well in style transfer. Pioneering works have been done especially in the image domain, so that transforming images of summer to winter, night to day and photos to certain painter's drawings can be done successfully. In this project, we focus on style transfer in the music domain.

There are several music datasets in the literature. However, lyrics and various instruments are mixed in some of them. To perform music genre transfer, these variants must be separated from each other. Therefore, due to time issues, we consider a music dataset containing only piano as an instrument. We work on transferring these symbolic music from source to target music genre domain. Hence, we use one of the state-of-the-art deep learning methods, CycleGAN \cite{zhu2017unpaired} which performs unpaired image-to-image translation using cycle consistent adversarial networks, as our baseline. After adapting CycleGAN framework to music domain, we improve our baseline by adding auxiliary discriminators and triplet loss. Thus, we perform music genre transfer so that melody of the source genre retains while note pitches change according to the target genre. To assess the performance of the model, we perform genre prediction on the transferred music and check the classification accuracy according to the target domain. To that end, we test several machine learning classification algorithms and choose Multi-Layer Perceptron classifier, which gives the best accuracy. Since it is a challenging task to evaluate transfer methods, we also run a subjective experiment.

%-------------------------------------------------------------------------

\section{Related Work}

\cite{brunner2018symbolic} introduces a method to perform music genre transfer on three major musical styles which are Jazz, Classical and Pop, by adapting CycleGAN model. They create and use a dataset consisting of MIDI (Musical Instrument Digital Interface) files. They introduce additional discriminators and classifiers to their CycleGAN-adapted approach. They train separate classifiers and transfer models for music genres. Their overall performance in both tasks is successful according to accuracies they obtained.

In their work, \cite{huang2018timbretron} presents a pipeline for musical timbre transfer using CycleGAN model. They use Constant-Q Transform\footnote{\url{https://en.wikipedia.org/wiki/Constant-Q_transform}} to convert audio files to audio spectrograms and WaveNet Synthesizer \cite{oord2016wavenet} to convert back to audio files. Their model captures the note pitches accurately and the sound quality in the outputs is high. Based on the results of human perceptual evaluations, they perform recognizable timbre transfer.

% https://cs230.stanford.edu/projects_fall_2018/reports/12445853.pdf (kullandıkları algolar, feature lar, katkıları, başarıları, 3 cümle)

Existing work on music genre classification includes \cite{cataltepe2007music}, where MIDI and audio features that are converted from MIDI files are classified using Linear Discriminant and k-Nearest Neighbor classifiers. The authors state that they obtain low accuracies with these classification methods. Therefore, they try combining these classification methods to increase the accuracy and in fact, they obtain accuracies as 99\% in Classic, 93\% in Jazz and 76\% in Pop genres.

%-------------------------------------------------------------------------

\section{The Approach}

Many deep learning methods for style transfer exist in the literature. These methods use extracted deep comprehensions of the domains to perform transfer. Especially deep generative models such as GANs are used for style transfer tasks. In this project, we mainly focus on performing style transfer in music genre domain by using GAN-based methods. We use and improve CycleGAN, one of the state-of-the-art deep learning methods, in style transfer.

In their work, \cite{zhu2017unpaired} introduces CycleGAN model to perform unpaired image-to-image translation by using cycle-consistent adversarial networks. The proposed model can perform image translation between horses and zebras, summer and winter, painting and photos, etc. Due to the success of the proposed model, CycleGAN has been used as a baseline in many studies. Hereby, we adapt this framework to music genre domain.

\begin{figure}[ht]
\vskip 0.1in
\begin{center}
\centerline{\includegraphics[width=\columnwidth]{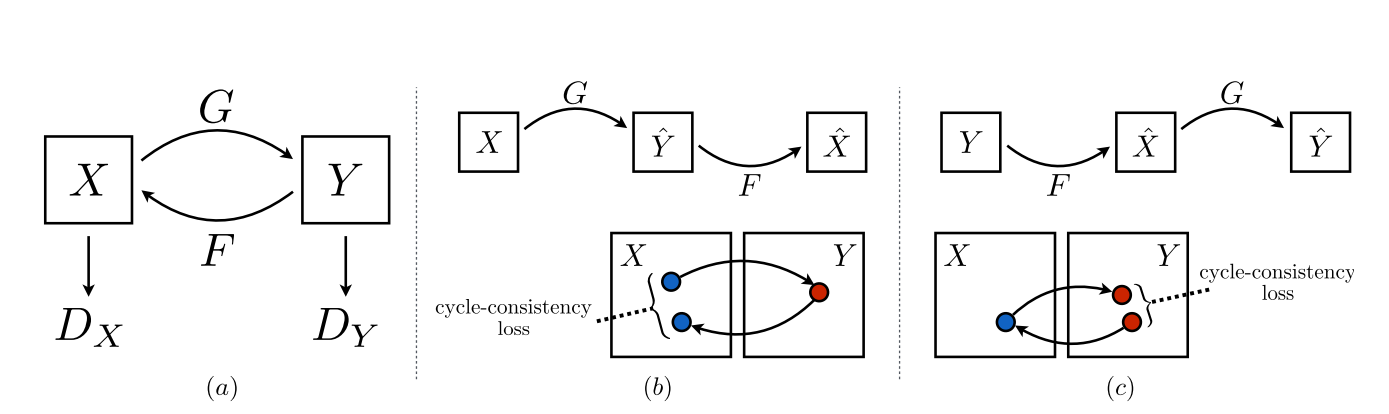}}
\caption{The model of CycleGAN. The sub-figure a shows cycles, sub-figures b and c show the detailed views of cycles.}
\label{fig:cyclegan_model}
\end{center}
\vskip -0.1in
\end{figure}

Figure \ref{fig:cyclegan_model} shows the model of CycleGAN. There are two domains X and Y, two generators G and F, two discriminators D\textsubscript{x} and D\textsubscript{y}. The main idea is when G generator maps X domain to Y domain, F generator maps the resulted Y domain back to X domain by using the discriminators. The cycle-consistency loss is calculated according to X domain and X domain obtained by F generator. By using this loss in training, model becomes more cycle-consistent so X domain and X domain obtained by F generator become identical.

\begin{figure}[ht]
\vskip 0.1in
\begin{center}
\centerline{\includegraphics[width=\columnwidth]{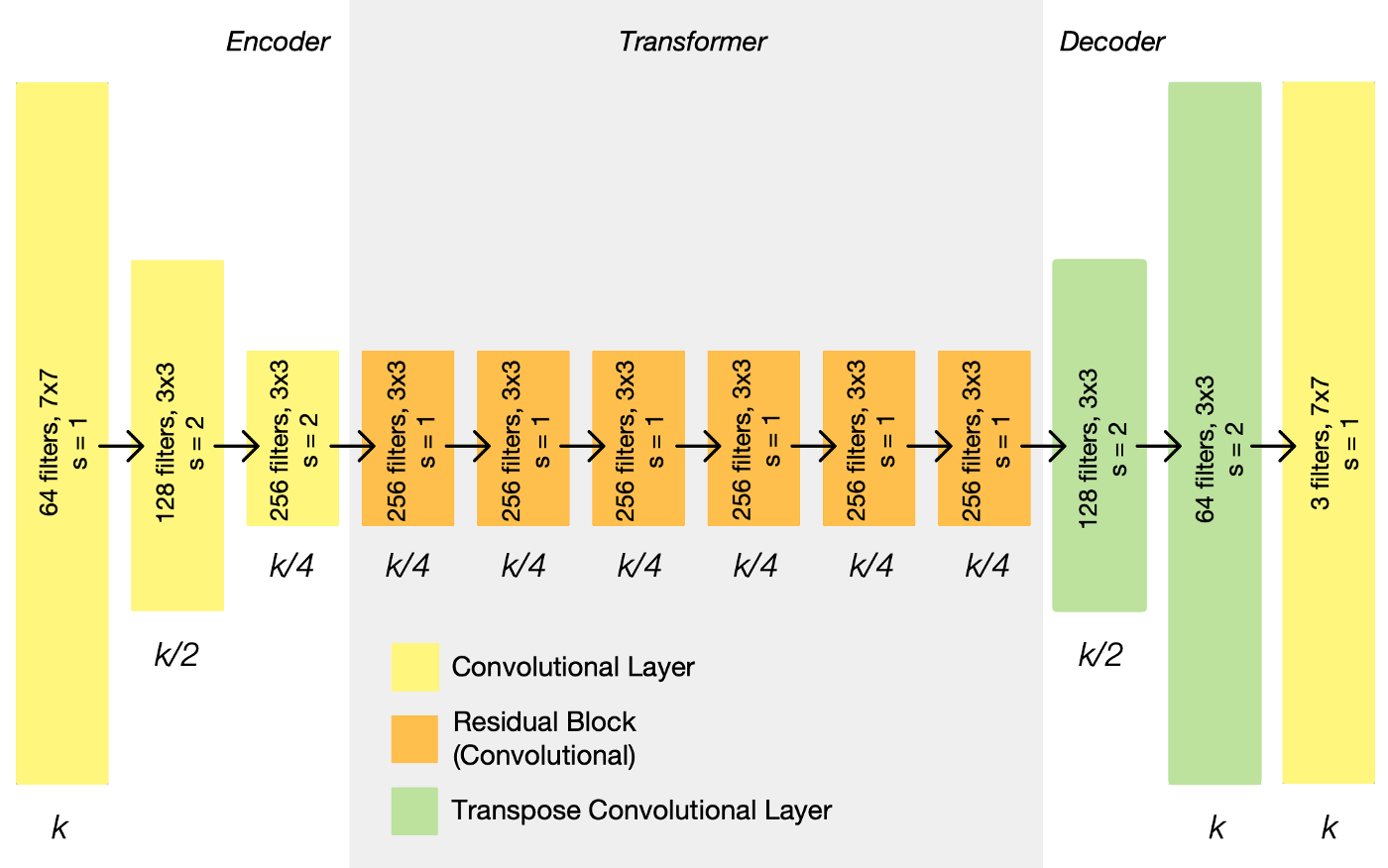}}
\caption{The architecture of CycleGAN}
\label{fig:cyclegan_arch}
\end{center}
\vskip -0.1in
\end{figure}

Figure \ref{fig:cyclegan_arch} shows the architecture of CycleGAN consisting of two GANs arranged in a cyclical fashion. This network contains two stride-2 convolutions, several residual blocks and two fractionally strided convolutions with stride $1/2$. In the transformer, there are 6 blocks for $128x128$ images, and 9 blocks for $256x256$ and higher resolution images. For the discriminator networks, there are $70x70$ PatchGANs which aim to classify whether $70x70$ overlapping image patches are real or fake. Such a patch-level discriminator architecture has fewer parameters than a full-image discriminator and can work on arbitrarily-sized images in a fully convolutional fashion.

To perform music genre transfer, we adapt this framework to music domain. We use the approach in \cite{huang2018timbretron}, where the authors use waveform images to train CycleGAN model. Therefore, we feed the MIDI files as images to CycleGAN model. Our dataset contains samples as matrices of size $64x84$ consisting of 0s and 1s, which represent the beats and notes. We convert these matrices into gray scale images. While converting, we do not crop or resize them because we need the outputs of the same size as the inputs, and we do not normalize them because our inputs already consist of 0s and 1s. Next, we change default activation function of CycleGAN from tanh to Sigmoid. By doing so, we get values between 0 and 1. Then, we convert the output images back to MIDI format by setting values smaller than a certain threshold to 0 and values greater than to 1. While converting, we use helper MIDI packages to save the images as MIDI format. Thus, we adapt CycleGAN framework to music domain and obtain MIDI outputs of the same size as the inputs.

CycleGAN model uses ResNet as its default generator network. To improve generations,  we experiment with well-known U-Net generator network. However, ResNet performs better in music genre transfer than U-Net according to our evaluation metric. Therefore, we continue to use ResNet in our model.

\begin{figure}[ht]
\vskip 0.1in
\begin{center}
\centerline{\includegraphics[width=\columnwidth]{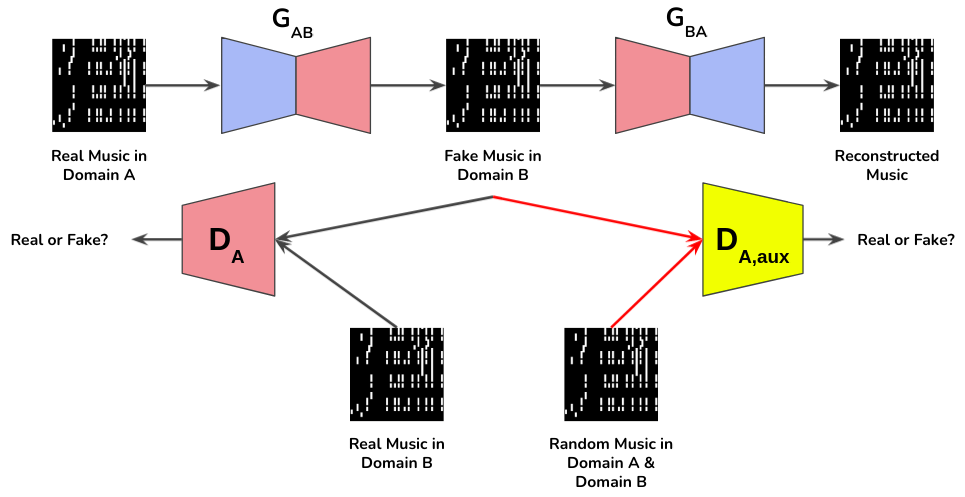}}
\caption{Our proposed auxiliary discriminator added model architecture. $D_{A,aux}$ represents the auxiliary discriminator for domain A to domain B transfer, which distinguishes transferred sample to domain B and a random sample from domain A and domain B. $D_{B,aux}$ follows the same steps for domain B to domain A transfer.}
\label{fig:aux_disc}
\end{center}
\vskip -0.1in
\end{figure}

In their work, \cite{brunner2018symbolic} performs symbolic music genre transfer by adapting CycleGAN model. To retain the content of input music in the transferred output music, they introduce additional discriminators. We follow the similar approach and add two auxiliary discriminators to our model for each domain. We take the output that is transferred to target domain and a random sample from the set of domain A and domain B. Auxiliary discriminators try to distinguish them and return the loss. By doing so, they force generators to learn higher level features. Figure \ref{fig:aux_disc} shows the auxiliary discriminator for domain A to domain B transfer. The other auxiliary discriminator is for domain B to A transfer, which basically follows the same steps as the figure shows.

CycleGAN model transfers sample from domain A to domain B, sample from domain B to domain A, and obtains fake samples, which have the following formulas.

\[ \hat{x}_{B} = G_{A\rightarrow B}(x_{A})  \]
\[ \hat{x}_{A} = G_{B\rightarrow A}(x_{B})  \]

By using these real and fake samples, standard discriminators try to distinguish original and transferred samples. Thus, following discriminator losses are used for standard GAN discriminators. Note that L2 norm is used for loss functions.

\[ L_{D_{A}} = \frac{1}{2} ( || D_{A}(x_{B}) || _2 + || D_{A}(\hat{x}_{B}) || _2 )  \]
\[ L_{D_{B}} = \frac{1}{2} ( || D_{B}(x_{A}) || _2 + || D_{B}(\hat{x}_{A}) || _2 )  \]

Unlike these GAN discriminators, auxiliary discriminators try to distinguish random samples from the set of domain A and B, and transferred samples. The key point here is to distinguish the transferred sample from not only the source domain but also from the target domain. So the model stick the music manifold while transferring, generates more realistic music, and transferred samples retain the content of the original samples. Thus, following auxiliary discriminator losses are used for each domain. Here, $M$ represents the mixed set of domain A and domain B, which are Jazz and Classic, and $x_{M}$ represents a randomly selected sample from that set.

\[ L_{D_{A,aux}} = \frac{1}{2} ( || D_{A,aux}(x_{M}) || _2 + || D_{A,aux}(\hat{x}_{B}) || _2 )  \]
\[ L_{D_{B,aux}} = \frac{1}{2} ( || D_{B,aux}(x_{M}) || _2 + || D_{B,aux}(\hat{x}_{A}) || _2 )  \]

For future experiments, we weight these auxiliary discriminator losses with $\gamma$. In our current experiment, we use $\gamma$ as 1. This value can be increased or decreased in next experiments, and its effect on the model can be observed. After weighting them, we feed standard and auxiliary discriminator losses to the model. At the end, we obtain the following formula for total discriminator loss.

\[ L_{D,total} = L_{D_A} +  L_{D_B} + \gamma (L_{D_{A,aux}} + L_{D_{B,aux}})  \]

To improve our model further, we add triplet loss. By using triplet loss, we minimize the distance between the transferred sample and the target domain while maximizing the distance between the transferred sample and the source domain. To do so, we take random samples from source and target domains while calculating the distances with transferred sample. Thus, we aim to achieve transfer results close to the target domain so that we can generate music with target genre style. We calculate two triplet losses for each domain as the following formulas show. Here, $x_{A,r}$ and $x_{B,r}$ denotes random samples from domain A and domain B respectively. In these formulas, we use default margin between these random samples, which is 1.

\[ L_{A, triplet} = max(|| \hat{x}_{B} - x_{B,r} ||^2 - || \hat{x}_{B} - x_{A,r} ||^2 + 1, 0)  \]
\[ L_{B, triplet} = max(|| \hat{x}_{A} - x_{A,r} ||^2 - || \hat{x}_{A} - x_{B,r} ||^2 + 1, 0)  \]

Then, we sum those triplet losses with other generator losses and feed this total generator loss to the model as the following formula shows. Here, $L_{G}$ denotes generator, cycle and identity losses.

\[ L_{G,total} = L_{G} + L_{A,triplet} + L_{B,triplet}  \]

We experiment triplet loss with and without auxiliary discriminators, and observe their effect on the music genre transfer task.

We evaluate the performance of our transfer models by using music genre classifier on the transferred music. First, we train several machine learning classification algorithms and choose the best music genre classifier by considering the test accuracy. Then, we check whether the obtained output from the model is classified as target music genre by using this classifier in testing and we present resulting accuracies. While evaluating the transfer models with music genre classifier, we rely on the prediction model. We choose the best transfer model according to the accuracies. After obtaining the best model, we run a subjective experiment for more accurate performance evaluation and present these results.

%-------------------------------------------------------------------------

\section{Experimental Settings}

\subsection{Dataset}

To perform music genre transfer, we need a dataset that contains note sequences or raw audio files to obtain audible outputs from the model. Since collecting such dataset takes too much effort, we need an existing dataset to perform our task. There are datasets that contain raw audio files in the literature, but it is hard to work with them. Since these audio files contain both vocals and numerous instruments, extracting audible features on given frequency spectrum are nearly impossible. Due to this limitation, we need to find dataset that contains MIDI files. A MIDI file is a symbolic music representation that holds note data like a sheet music. MIDI format is created for communication between electrical instruments and computers or synthesizers. MIDI files do not contain any sound or audio wave. They just indicate when to press a note and when to release a note. As a result of this representation, music style transfer can be done without any influence of wave data.

\begin{figure}[ht]
\vskip 0.1in
\begin{center}
\centerline{\includegraphics[width=\columnwidth]{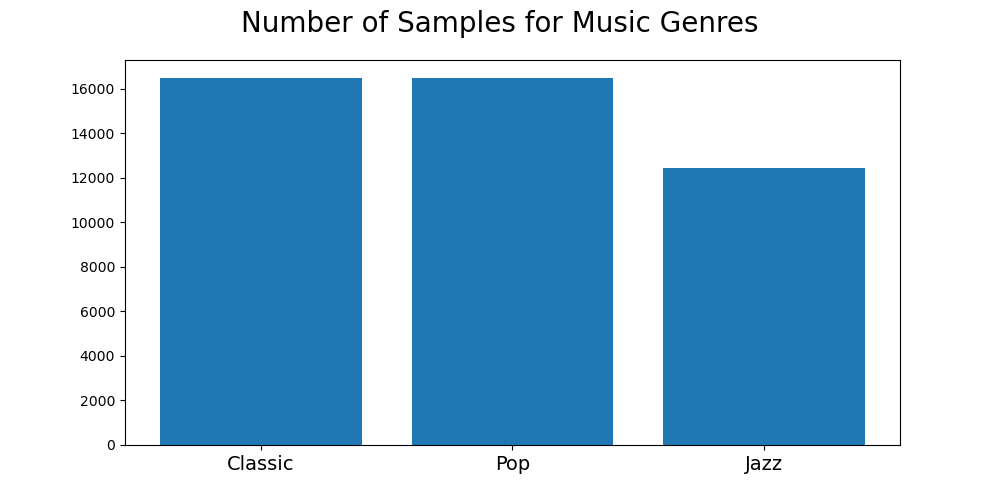}}
\caption{Number of samples for each music genre of the dataset proposed in \cite{brunner2018symbolic}}
\label{fig:genre_cyclegan}
\end{center}
\vskip -0.1in
\end{figure}

In their work, \cite{brunner2018symbolic} presents a MIDI dataset that contains both genre information and cleaned MIDI files. The authors downloaded MIDI files from numerous sites and compiled them into three genres including Jazz, Classic and Pop. Since MIDI files were compiled from numerous sources, they did some preprocessing. They initially filtered out MIDI files that first beat did not start at 0. Then, they filtered out MIDI files where the time signatures changed while playing the song, or time signature was not $4/4$. These MIDI files were then converted into numpy arrays for easier training processes. The obtained dataset after preprocessing have the distribution over genres as in Figure \ref{fig:genre_cyclegan}. This figure contains samples from both training and test data. We use this dataset in style transfer and prediction tasks. However, we only consider Jazz and Classic genres due to time and compute power limitations.

%We train our models with music in the MIDI format, since these MIDI files are in numpy arrays, we can simply feed these matrices to our network.

%-------------------------------------------------------------------------

\subsection{Music Genre Prediction}

\subsubsection{Naive Bayes Algorithm}

Naive Bayes classifier is based on conditionally independence assumption and classifies each element according to the highest probability. It can achieve successful results with little training data. In our dataset, this algorithm achieved 78.95\% accuracy. The reason of this outcome maybe that the features do not come from normal distribution.

\subsubsection{K-Nearest Neighbors Algorithm}
\begin{figure}[ht]
\vskip 0.1in
\begin{center}
\centerline{\includegraphics[width=\columnwidth]{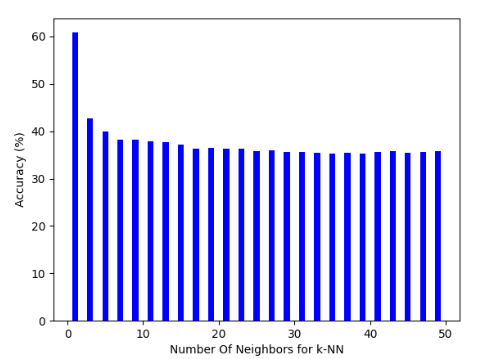}}
\caption{Test accuracies according to different k values in k-NN algorithm}
\label{fig:knn}
\end{center}
\vskip -0.1in
\end{figure}

K-Nearest Neighbors (k-NN) is a non-parametric classification algorithm that relies on distances between samples and make classifications according to the closest ones. Therefore, the success of the algorithm is related to the fact that the assumption on closer samples are similar to each other is valid enough. For the algorithm to make accurate predictions with this assumption, it is important to tune hyperparameter k, which indicates the number of neighbors considered to classify the samples in the algorithm. Therefore, we tuned this hyperparameter by using values between 1-50 range. Figure \ref{fig:knn} shows the obtained test accuracies according to these different k values. The fact that the best accuracy obtained is when k is 1 maybe related with the clutter of our dataset. We achieved the best accuracy as 61.05\% with 1 as k value and the Euclidean distance as the distance metric. Since k-NN is a non-parametric algorithm, it has limitations. Therefore, the obtained accuracy is not high enough.

\subsubsection{Random Forest Algorithm}

\begin{figure}[ht]
\vskip 0.1in 
\begin{center}
\centerline{\includegraphics[width=\columnwidth]{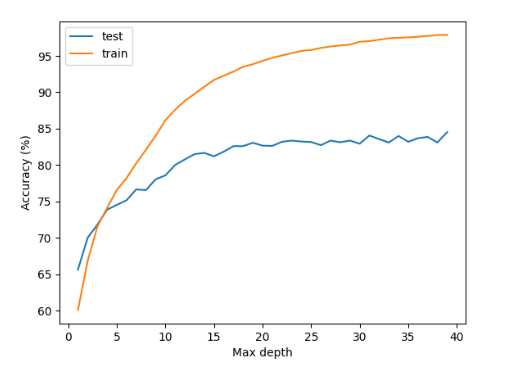}}
\caption{Train and test accuracies according to different max depths in Random Forest algorithm}
\label{fig:rfa}
\end{center}
\vskip -0.1in
\end{figure}

Random Forest is an ensemble learning method of decision trees. It is a supervised machine learning algorithm used for both classification and regression tasks. It works by creating multiple decision trees, combining them together for more accurate and stable decisions. While applying this algorithm, we tuned max depth hyperparameter by using values between 1-40 range. Figure \ref{fig:rfa} shows the obtained test accuracies according to these different max depth values. Although train accuracy increases as the max depth increases, test accuracy does not increase significantly with max depths greater than 20. We obtained the best accuracy as 84.32\% with this algorithm. Since Random Forest algorithm works well with high dimensional data, robust to overfitting and outliers, it obtained accuracy better than Naive Bayes and k-NN algorithms.

\subsubsection{Multi-layer Perceptron Algorithm}

Multi-layer Perceptron (MLP) is a deep artificial neural network that is composed of several perceptrons. MLP classifiers consist of an input layer, an output layer and an arbitrary number of hidden layers between them. They learn the patterns and dependencies between fed inputs and outputs by adjusting the weights and biases, and minimizing the obtained error. There are several hyperparameters to tune for this algorithm. We tuned as many parameters as our compute power allowed, which are number of hidden layers, number of neurons in each hidden layer, activation function, solver for weight optimization, learning rate schedule and maximum number of iterations. We obtained the best test accuracy as \%87.72 with 1 hidden layer, 1000 hidden neurons, ReLU activation function, Adam optimizer, adaptive learning rate and 1000 iterations. This model achieves 90\% accuracy in Classic genre and 83\% accuracy in Jazz genre. Since MLP algorithm can learn non-linear functions and has high generalization capacity, it obtained better results than other algorithms. Therefore, we use this model to evaluate the performance of the style transfer models.

\begin{table}[ht]
\caption{Best test accuracies obtained with different algorithms}
\vskip 0.1in
\begin{center}
\begin{small}
\begin{sc}
\begin{tabular}{lcccr}
\toprule
Algorithm & Test Accuracy (\%) \\
\midrule
Naive Bayes & 78.95 \\
k-Nearest Neighbors & 61.05 \\
Random Forest & 84.32 \\
\textbf{Multi-layer Perceptron} & \textbf{87.72} \\
\bottomrule
\label{table:acc}
\end{tabular}
\end{sc}
\end{small}
\end{center}
\vskip -0.1in
\end{table}

%-------------------------------------------------------------------------

\section{Experimental Results}

We work on performing music genre transfer by adjusting the CycleGAN framework, which is widely used on image-to-image style transfer task, to music domain. Then we add auxiliary discriminators and triplet loss in order to improve the baseline model. After experimental setups, we train the models using NVIDIA GTX 1080ti GPU, which lasts about 16 hours. Then, we test and evaluate our models by using our music genre classifier, and obtain acccuracies as evaluation metric. The evaluation of the style transfer methods is challenging, since it is a highly subjective measure. Therefore, we run a subjective experiment on our best style transfer model selected with best accuracy. In this section, we present obtained accuracies and images representing MIDI formats on test data of our style transfer models, and user study results of our best model. The audio samples according to our proposed models can be found here\footnote{\url{https://youtube.com/playlist?list=PLjt5tUIGgtC6TLn8TG_al1Obzxz14rVVS}}.

%-------------------------------------------------------------------------

\subsection{Generator Networks}
After adapting CycleGAN framework to music domain, we perform Jazz to Classic transfer task with the baseline model. To increase the success of the generator, we try ResNet with 9 blocks, which is the default generator network of CycleGAN, and U-Net with 128 blocks generator networks. As a result, ResNet obtains 56\% accuracy while U-Net obtains 10\% accuracy. Due to the gap between these accuracy results, we do not train the model with U-Net generator network on Classic to Jazz transfer task for comparison. We continue to use ResNet generator network in our baseline. As Table \ref{table:transfer-acc} shows our baseline model obtains 56\% accuracy in Jazz to Classic transfer task and 38\% accuracy in Classic to Jazz transfer task. As Figure \ref{transfer-img} shows our baseline model can make minor random changes and remove some notes.

%-------------------------------------------------------------------------

\subsection{Auxiliary Discriminators}
To retain the structure of input in the transferred output, we add auxiliary discriminators to our model. These discriminators try to distinguish transferred output and a random sample from mixed set of domains. By doing so, we force the generators to learn higher-level features, stick the music manifold and produce more realistic music. As Table \ref{table:transfer-acc} shows adding auxiliary discriminators decreases the performance of the baseline model, although this approach has a positive effect in the model proposed in \cite{brunner2018symbolic}. As Figure \ref{transfer-img} shows although the model manages to make changes, it has lower effect than baseline model on transfer.

%-------------------------------------------------------------------------

\subsection{Triplet Loss}
For closer generations to target domain, we add triplet loss to the baseline model. By using triplet loss, we maximize the distance between the generated samples and their source domain, while minimizing the distance between their target domain. In fact, using this loss helps us to improve the baseline performance more than 10 points in Jazz to Classic transfer task as shown in Table \ref{table:transfer-acc}. However, it can not improve the baseline performance in Classic to Jazz transfer task. The reason of this outcome maybe that this task is more challenging than the Jazz to Classic transfer task. As Figure \ref{transfer-img} shows this model removes, adds and even concatenates some notes.

%-------------------------------------------------------------------------

\subsection{Auxiliary Discriminators \& Triplet Loss}
\begin{figure*}
\centering
\setlength\tabcolsep{0pt} % default value: 6pt
\hspace*{-1.0cm}%
\begin{tabular}{ccccc}
\hspace*{0.0cm}Input & \hspace*{0.25cm}CycleGAN & \hspace*{0.25cm}D\textsubscript{aux} & \hspace*{0.25cm}Triplet Loss & \hspace*{0.25cm}D\textsubscript{aux}\& Triplet Loss \\

%%%% 1st row
\includegraphics[width=0.23\textwidth, keepaspectratio,]{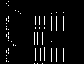} & 
\includegraphics[width=0.23\textwidth, keepaspectratio,]{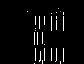}&
\includegraphics[width=0.23\textwidth, keepaspectratio,]{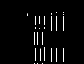}&
\includegraphics[width=0.23\textwidth, keepaspectratio,]{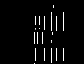}&
\includegraphics[width=0.23\textwidth, keepaspectratio,]{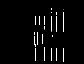}\\

%%%% 2nd row
\includegraphics[width=0.23\textwidth, keepaspectratio,]{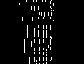} & 
\includegraphics[width=0.23\textwidth, keepaspectratio,]{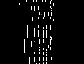}&
\includegraphics[width=0.23\textwidth, keepaspectratio,]{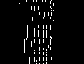}&
\includegraphics[width=0.23\textwidth, keepaspectratio,]{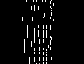}&
\includegraphics[width=0.23\textwidth,  keepaspectratio,]{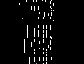}\\

\end{tabular}
\caption{Example results of Jazz to Classic transfer task according to applied experiments. The rows represent samples and model results, where \nth{1} column represents the inputs. The remaining columns show the outputs of baseline, auxiliary discriminators added, triplet loss added and auxiliary discriminators \& triplet loss added models respectively.}
\label{transfer-img}
\end{figure*}

We combine our contributions to observe their effect in the model. In fact, their combination obtains the best results as Table \ref{table:transfer-acc} shows. It increases the baseline accuracy up to 14 points in Jazz to Classic transfer task and 1 point in Classic to Jazz transfer task. Although using auxiliary discriminators alone in the baseline model decreases the performance, combining them with triplet loss improves the baseline model and triplet loss alone added model. This outcome shows the positive effect of the approaches to each other for the model performance. Even the performance of this model can be improved by increasing or decreasing the weight of the auxiliary discriminator losses. This model succeeds by combining the changes made by the two models as Figure \ref{transfer-img} shows.

\begin{table}[ht]
\caption{Results of proposed models according to translation tasks}
\vskip 0.1in
\begin{center}
\begin{small}
\begin{sc}
\begin{tabular}{lcccr}
\toprule
Model & J2C Acc (\%) & C2J Acc (\%) \\
\midrule
Baseline & 56 & 38 \\
+ D\textsubscript{aux} & 53 & 35 \\
+ Triplet Loss & 69.3 & 37.7 \\
\textbf{+ D\textsubscript{aux} + Triplet Loss} & \textbf{69.4} & \textbf{39.3} \\
\bottomrule
\label{table:transfer-acc}
\end{tabular}
\end{sc}
\end{small}
\end{center}
\vskip -0.1in
\end{table}

%%%%% 1
%%%%% 1

\subsection{Subjective Experiment}

In this project, we work on style transfer, which is a challenging task to evaluate the performance. Throughout the project, we rely on our music genre classifier and evaluate the style transfer models by using this classifier. However, this task requires a subjective experiment where users asses the performance of the proposed model. After selecting the best style transfer model with accuracies, we perform a user study with this selected model. After examining the audio results of our baseline model, we see that there are no significant changes in the transferred outputs. Also, people are tend not to join long studies. In the light of these outcomes, we only consider the results of our proposed model in the user study. 23 people participated to our study.

Our user study consists of 2 parts. The first part consists of six questions, where a transferred music from test data is given in each question to choose its genre. In these question, we ask users to select genres of given music from Classic, Jazz and other genres. In Table \ref{table:partA}, we see the accuracies of six different generated outputs obtained from the user study. Average accuracy of Jazz to Classic transfer task, which shows high results among the two models, is 60.8\% and Classic to Jazz transfer task is 56.5\%. These results from the user study are close to the results that we expected. Therefore, the overall performance of our transfer models are successful.

\begin{table}[ht]
\caption{Results of the proposed models in user study}
\vskip 0.1in
\begin{center}
\begin{small}
\begin{sc}
\begin{tabular}{lcccc}
\toprule
Task & Jazz & Classic & Other & Acc (\%) \\
\midrule
\multirow{3}{*}{Jazz2Classic} & 1 & \textbf{22} & 0 & 95.6 \\
& 10 & \textbf{11} & 2 & 47.8 \\
& 9 & \textbf{9} & 5 & 39.1 \\
\hline
\multirow{3}{*}{Classic2Jazz} & \textbf{15} & 5 & 3 & 65.2 \\
& \textbf{13} & 6 & 4 & 56.5 \\
& \textbf{11} & 5 & 7 & 47.8 \\
\bottomrule
\label{table:partA}
\end{tabular}
\end{sc}
\end{small}
\end{center}
\vskip -0.1in
\end{table}

The second part consists of two questions, where a test sample and its transferred output are given in each question to score the melody conservation. In our proposed model, we try to generate music that retains the structure of the source genre to stay on the music manifold. Therefore, in these questions, we ask users to rate the melody conservation between original samples and transferred outputs with scores 1 to 5. 1 means the melody conservation between input and output is poor while 5 means it is strong.

In Table \ref{table:partB}, the average rate of melody conservation obtained according to the values between 1 to 5 taken from the user study are shown. According to the average values obtained, the melody conservation between inputs and outputs are nearly strong in both tasks.

\begin{table}[ht]
\caption{Melody conservation scores in user study}
\vskip 0.1in
\begin{center}
\begin{small}
\begin{sc}
\begin{tabular}{lcccccc}
\toprule
Task & 1 & 2 & 3 & 4 & 5 & Avg \\
\midrule
Jazz2Classic & 2 & 0 & 5 & 13 & 3 & 3.65 \\
Classic2Jazz & 2 & 3 & 5 & 6 & 7 & 3.57 \\
\bottomrule
\label{table:partB}
\end{tabular}
\end{sc}
\end{small}
\end{center}
\vskip -0.1in
\end{table}

%-------------------------------------------------------------------------

\section{Limitations}
There are some limitations in our approaches, which are as follows.

\begin{itemize}
    \item There are no diverse works in music genre transfer topic in the literature. Existing ones generally do not share a common approach for solving musical style transfer problem. Therefore, we use a baseline model that performs image-to-image translation task and adapt it to music domain from scratch. 
    \item We perform music genre transfer on two music genres. In terms of variation, we need to perform style transfer in different music genres. Further, we need a dataset where we can separate the instruments from the songs. Therefore, we can transfer genres in songs instead of just symbolic music.
    \item We have limited time and compute power, which are great issues for training deep learning models.
    \item Evaluating the performance of transfer methods is dependant to the subjective measure. At first, we find the best transfer model with genre classifier and then, we evaluate the best model with user study. Although we rely on people's musical knowledge in the user study, this may not always be the case.
    \item Our method is not successful enough for Classic to Jazz translation task. Also, the prediction model performs poorly on classifying Jazz genre. Therefore, Classic to Jazz translation task can be more challenging than Jazz to Classic.
\end{itemize}

%-------------------------------------------------------------------------

\section{Conclusions}
% TODO Oguz: This section summarizes all your project work, focusing on the key results you obtained. You may also suggest possible directions for future work

In this project, we present approaches to music genre transfer. To evaluate these unsupervised learning methods, we use a music genre classifier for measurable results. For this intent, we train four different classification algorithms using the training sets of our dataset. Best test accuracy for this task is obtained as 87.72\% with Multi-Layer Perceptron classifier. In the future, this classifier can be changed with better one, for instance with CNN-based classification methods, to obtain more accurate results as style transfer performance measure. Thus, we can eliminate wrong evaluations especially in Classic to Jazz translation task.

In addition to style transfer evaluation with genre classifier, we run a subjective experiment to support our results. According to the user study, overall performance of the transfer models are good and we manage to conserve the melody of music. In the future, we can reach more experienced users for these evaluations to achieve more consistent results.

To perform music genre transfer, we use CycleGAN as our baseline to enable style transfer between two domains. Our methods benefit from our strong baseline. In \cite{brunner2018symbolic}, they propose additional discriminators to regularize the generators. These discriminators enable generators to preserve the structure of the input and perform style transfer between domains. In our experiments, we conclude that these auxiliary discriminators decrease accuracy of the baseline model.

Our proposed model with triplet loss perform significantly better in Jazz to Classic style transfer task. By adding both auxiliary discriminators and triplet loss to our baseline, we observe small increases in both Jazz to Classic and Classic to Jazz transfer tasks. However, this model obtains the best accuracy among our experiments. In the future, weight of auxiliary discriminators can be tuned and the better results can be obtained.

For future direction, we can improve our model further by applying new approaches. After obtaining a classification model that achieves 98-99\% test accuracy, we can use it to feed classification loss in training. Later, we can use different GAN models such as DiscoGAN model proposed in \cite{kim2017learning}. Lastly, we can apply our model to new music genres.

%In future direction, this method can be extended or changed to achieve better classification results. In addition to these changes, we could not train with different gamma values in auxiliary discriminators. Optimizing this gamma value may increase or decrease classification results. After adding these auxiliary discriminators to our proposed triplet loss method, we observe minimal increase in accuracy compared to our proposed triplet loss method. 

%-------------------------------------------------------------------------

\newpage

%-------------------------------------------------------------------------

\bibliography{report}
\bibliographystyle{icml2021}

\end{document}